\documentclass[11pt]{article}
\newcommand{\nc}{\newcommand}
\nc{\bib}{\bibitem}
\nc{\al}{\alpha}
\nc{\g}{\gamma}
\nc{\G}{\Gamma}
\nc{\D}{\Delta}
\nc{\eps}{\epsilon}
\nc{\la}{\lambda}
\nc{\La}{\Lambda}
\nc{\var}{\varphi}
\nc{\pa}{\partial}
\nc{\nn}{\nonumber \\ }
\nc{\hf}{\frac{1}{2}}         
\nc{\dz}{\frac{dz}{2\pi i}}
\nc{\bin}[2]{\left (\begin{array}{c} {#1}\\ {#2} \end{array}\right )}
\nc{\ben}{\begin{equation}}
\nc{\een}{\end{equation}}
\nc{\bea}{\begin{eqnarray}}
\nc{\eea}{\end{eqnarray}}
\nc{\bra}[1]{\langle {#1}|}
\nc{\ket}[1]{|{#1}\rangle}
\nc{\C}{\mbox{\hspace{1.24mm}\rule{0.2mm}{2.5mm}\hspace{-2.7mm} C}}
\nc{\Nat}{\mbox{\hspace{.04mm}\rule{0.2mm}{2.8mm}\hspace{-1.5mm} N}}
\nc{\NP}[1]{Nucl.\ Phys.\ {\bf #1}}
\nc{\PL}[1]{Phys.\ Lett.\ {\bf #1}}
\nc{\CMP}[1]{Commun.\ Math.\ Phys.\ {\bf #1}}
\nc{\PR}[1]{Phys.\ Rev.\ {\bf #1}}
\nc{\PRL}[1]{Phys.\ Rev.\ Lett.\ {\bf #1}}
\nc{\PTP}[1]{Prog.\ Theor.\ Phys.\ {\bf #1}}
\nc{\PTPS}[1]{Prog.\ Theor.\ Phys.\ Suppl.\ {\bf #1}}
\nc{\MPL}[1]{Mod.\ Phys.\ Lett.\ {\bf #1}}
\nc{\IJMP}[1]{Int.\ Jour.\ Mod.\ Phys.\ {\bf #1}}
\nc{\IM}[1]{Invent.\ Math.\ {\bf #1}}
\nc{\SJNP}[1]{Sov. J. Nucl. Phys.\ {\bf #1}}
\nc{\JHEP}[1]{J.\ High\ Energy Phys.\ {\bf #1}}

\def\vvdots{\mathinner{\mkern1mu\raise1pt\vbox{\kern7pt\hbox{.}}\mkern2mu
 \raise4pt\hbox{.}\mkern2mu\raise7pt\hbox{.}\mkern1mu}}

\begin{document}

\topmargin -5mm
\oddsidemargin 5mm

\begin{titlepage}
\setcounter{page}{0}

\vspace{8mm}
\begin{center}
{\huge Purely affine elementary $su(N)$ fusions}

\vspace{15mm}
{\large J{\o}rgen Rasmussen}\footnote{rasmussj@cs.uleth.ca; address after
1 December 2001: CRM, Universit\'e de Montr\'eal, Case postale 6128, 
succursale centre-ville, Montr\'eal, Qu\'ebec, Canada H3C 3J7} and 
{\large Mark A. Walton}\footnote{walton@uleth.ca}
\\[.2cm]
{\em Physics Department, University of Lethbridge,
Lethbridge, Alberta, Canada T1K 3M4}

\end{center}

\vspace{10mm}
\centerline{{\bf{Abstract}}}
\vskip.4cm
\noindent
We consider three-point couplings in simple Lie algebras -- 
singlets in triple tensor products of their 
integrable highest weight representations. A coupling can be 
expressed as a linear combination of products of finitely many 
elementary couplings. 
This carries over to affine fusion, the fusion of 
Wess-Zumino-Witten conformal field theories, where the 
expressions are in terms of elementary fusions. 
In the case of $su(4)$ it has been observed 
that there is a purely affine elementary fusion, i.e., an
elementary fusion that
is not an elementary coupling. In this note we show by construction 
that there is at least one purely affine elementary fusion 
associated to every $su(N>3)$.  
\end{titlepage}
\newpage
\renewcommand{\thefootnote}{\arabic{footnote}}
\setcounter{footnote}{0}

\section{Introduction}

It is well-known that the set of elementary (three-point) couplings
of any simple Lie algebra is finite. It is natural to expect that 
three-point fusions may also be expressed in terms of 
non-negative linear combinations of products 
of finitely many elementary fusions (EFs). A fusion may be 
labelled by the weights $(\la,\mu,\nu)$ of the associated
coupling, and its threshold level $t$: $(\la,\mu,\nu)_t$.
It has been conjectured \cite{CMW} that fusion multiplicities
are uniquely determined from the tensor product multiplicities 
$T_{\la,\mu,\nu}$ and the associated multi-set of minimum levels $\{t\}$ 
at which the various couplings first appear.\footnote{Let us review 
here the status of this conjecture. Modulo rigour, it follows from the 
Gepner-Witten depth rule \cite{GW}. A rigorous proof would follow from a 
refinement of the depth rule conjectured in \cite{kmsw}. A 
version of this formula was independently proved in \cite{FZ} in the 
context of vertex operator algebras (VOAs). 
The only thing missing at this point  
is the proof that the fusion rules of affine VOAs, as calculated 
in \cite{FZ}, 
are identical to those given by the Verlinde formula.}
Therefore, to the triplet $(\la,\mu,\nu)$ there correspond
$T_{\la,\mu,\nu}$ distinct couplings, hence $T_{\la,\mu,\nu}$ values
of $t$, one for each distinct coupling. These values are called
threshold levels. The threshold levels associated to two different
couplings may be identical. 

All elementary
couplings must be EFs when assigned the associated threshold
levels. However, there may in general be more EFs than elementary
couplings. The extra EFs will be called purely affine EFs. The only known
example is in $su(4)$ \cite{BKMW,RW5}, and it reads
\ben
 (\La^1+\La^3,\La^1+\La^3,\La^1+\La^3)_2
\label{su4}
\een
Here $\{\La^i\}_{i=1,...,r}$ is the set of fundamental weights,
and $r$ is the rank of the Lie algebra. The tensor product multiplicity
is two:
\ben
 T_{\La^1+\La^3,\La^1+\La^3,\La^1+\La^3}=2
\een
and the set of threshold levels is $\{2,3\}$.

The objective of the present work is the extension of this result to 
higher rank. Thus, we shall prove that for {\em every} $su(r+1)$, $r\geq3$,
\ben
 (\La^1+\La^r,\La^1+\La^r,\La^1+\La^r)_2
\label{sun}
\een
is a purely affine EF. As for $su(4)$, the tensor product multiplicity
is two: 
\ben
 T_{\La^1+\La^r,\La^1+\La^r,\La^1+\La^r}=2
\een
and the set of threshold levels is $\{2,3\}$.

A knowledge of EFs is important, for example, when constructing
three-point functions in Wess-Zumino-Witten theories along the lines
discussed in \cite{RW5}. There and here we use a formalism where couplings
are represented by polynomials, and the threshold levels correspond
to particular properties of those. We construct the polynomial for 
(\ref{sun}) in sect. 3 below. 

The existence of (\ref{sun}) as a purely affine elementary fusion 
may also be proved just by counting fusion multiplicities, as we 
will show. The corresponding polynomial contains more information, 
however, and in the process of constructing it, we will explain 
why purely affine elementary fusions can appear. On the other hand, 
the simpler counting argument is useful in that it can be 
much more easily adapted to algebras besides $su(N)$, as we will indicate.   

\section{Background}

Here we shall provide the necessary background and 
fix our notation, by discussing certain results on
realizations of Lie algebras, their highest weight modules, tensor products,
affine fusion, threshold levels, and elementary couplings.

\subsection{Differential operators and polynomial realizations}

A general and convenient differential operator realization of any simple
Lie algebra (in particular $su(r+1)$)  was provided in \cite{Ras1,PRY}.
It is given in terms of the flag variables (or triangular coordinates)
of the Lie algebra -- there is one independent parameter $x^\al$ for every
positive root $\al>0$. Let $\la$ be the highest weight of the representation,
then according to \cite{Ras1,PRY},
\bea
 E_\al(x,\pa)&=&\sum_{\beta>0}V_\al^\beta(x)\pa_\beta\nn
 H_i(x,\pa,\la)&=&\sum_{\beta>0}V_i^\beta(x)\pa_\beta+\la_i\nn
 F_\al(x,\pa,\la)&=&\sum_{\beta>0}V_{-\al}^\beta(x)\pa_\beta+
  \sum_{j=1}^rP_{-\al}^j(x)\la_j
\label{diff}
\eea
is a differential operator realization of the Lie algebra. $V$ and $P$
are polynomials in the flag variables $x$ provided in \cite{Ras1,PRY}. 
Here we only need to know explicitly
\bea
 E_\theta(x,\pa)&=&\pa_\theta\nn
 P_{-\al}^j(x)&=&\left[e^{-C(x)}\right]_{-\al}^j
\label{EP}
\eea
where $C_a^b(x)=-\sum_{\beta>0}x^\beta{f_{\beta,a}}^b$ with ${f_{a,b}}^c$
being a structure constant. $\theta$ is the highest root.
If ${\rm wt}(x^\al\cdots x^\beta)=\al+...+\beta$
and wt is linear, it follows that $P$ is homogeneous in its weight,
and that it is ${\rm wt}(P_{-\al}^j(x))=\al$.

One further utilization of the flag variables is to represent states in
highest weight modules $M_\la$ \cite{PRY}, or tensor products thereof 
\cite{Ras2,Ras3}, as polynomials. For example, the polynomial (or wave 
function) associated to the state
\ben
 F_\al\cdots F_\beta\ket{\la}
\een
is
\ben
 b_{\al,...,\beta}(x,\la)=F_\al(x,\pa,\la)\cdots F_\beta(x,\pa,\la)1
\een
Likewise, the polynomial associated to
\ben
 F_{\al^{(1)}}\cdots F_{\beta^{(1)}}\ket{\la^{(1)}}\otimes\cdots \otimes 
  F_{\al^{(n)}}\cdots F_{\beta^{(n)}}\ket{\la^{(n)}}
\label{prod}
\een
is the following ordinary product of polynomials in the $n$ sets of flag 
variables $x_1,...,x_n$:
\ben 
 b_{\al^{(1)},...,\beta^{(1)}}(x_1,\la^{(1)})\cdots 
  b_{\al^{(n)},...,\beta^{(n)}}(x_n,\la^{(n)})
\een
We are interested in the linear combinations of states like (\ref{prod})
that correspond to the singlet, and we shall focus mainly on the case $n=3$.
That corresponds to considering three-point couplings (to the singlet).

In \cite{RW5}, we showed how the polynomial description above is ideally
suited to the implementation of the Gepner-Witten depth rule \cite{GW}
of affine fusion.
The rule encodes the level-dependence. Incorporating the idea of threshold
levels, the key point is the following. A polynomial associated to a
(three-point) coupling may be assigned the threshold level given by the 
highest power of $x^\theta$, using $x^\theta_l=x^\theta$ for all
$l=1,2,3$. That follows essentially from the differential operator
realization of $E_\theta$ (\ref{diff}).

Certain relations (or syzygies) among different couplings and
their decompositions into elementary couplings, complicate the
determination of the multi-set of threshold levels. We will have more to
say about that below.

\subsection{Elementary couplings and threshold levels}

Let $E$ denote a generic element in the set ${\cal E}$ of elementary 
(three-point) couplings. The associated elementary polynomial is
$R^E(x_1,x_2,x_3)$. Consider a decomposition of the coupling
$(\la,\mu,\nu)$ on ${\cal E}$:
\ben
 (\la,\mu,\nu)=\prod_{E\in{\cal E}}E^{\times c_E}
\label{dec}
\een
where $c_E$ are non-negative integers. We indicate a multiplication of 
couplings by $\times$, e.g., $E\times E'$ and $E^{\times2}=E\times E$. 
The polynomial associated to (\ref{dec}) enjoys the factorization
\ben
 R^{\la,\mu,\nu}(x_1,x_2,x_3)=\prod_{E\in{\cal E}}\left(R^E(x_1,x_2,x_3)
  \right)^{c_E}
\label{fac}
\een
It follows that we can assign the threshold level
\ben
 t(R^{\la,\mu,\nu})=\sum_{E\in{\cal E}}c_E\,\, t(R^E)
\label{tR}
\een
to (\ref{fac}). Actually, a similar result holds even if the factorization
(\ref{fac}) is not on ${\cal E}$.

Decompositions like (\ref{dec}) are in general not unique. That results
in certain algebraic relations among the elementary couplings (and so
among the elementary polynomials) that must be taken
into account. These relations are sometimes called syzygies. They can
be implemented by excluding certain combinations of the elementary 
couplings (or products of the elementary polynomials). The resulting
number of allowed decompositions (\ref{dec}) is the associated
tensor product multiplicity. Here we do not need any
detailed knowledge of that procedure.

Having obtained the $T_{\la,\mu,\nu}$ linearly independent polynomials
(\ref{fac}) associated to the triplet $(\la,\mu,\nu)$,
$\{R^{\la,\mu,\nu}_s(x_1,x_2,x_3)\}_{s=1,...,T_{\la,\mu,\nu}}$,
we are still faced with the problem of determining the multi-set of
threshold levels. Naively, one could just ``measure'' the threshold
levels of each of the polynomials $R^{\la,\mu,\nu}_s$, but that would in 
general produce a multi-set of values slightly greater than the actual
multi-set of threshold levels. The reason is found in the fact that
$\{R^{\la,\mu,\nu}_s\}$ is merely a convenient basis for the description
of the {\em tensor product}.
When discussing fusion, i.e., when also determining the multi-set of
threshold levels, one might need a different basis. Any such basis will
consist of linear combinations of the original basis elements
$R^{\la,\mu,\nu}_s$. Now, it is obvious that the threshold level of the sum 
of polynomials in general will be equal to the maximum value
of the individual threshold levels. Nevertheless, it may occur that a
linear combination of polynomials will have
a {\em smaller} maximum power of $x^\theta$ than {\em any} of the
participating polynomials. That is precisely what 
we shall encounter below.

A subset of the elementary three-point couplings
is governed by the two-point couplings, since a two-point coupling is
a three-point coupling with one of the three weights being zero.
The set of elementary two-point couplings is easily described.
It consists of the couplings $(\La^i,{\La^i}^+,0)$, $i=1,...,r$,
and permutations thereof. Thus, the number of independent,
elementary two-point couplings is $3r$.
 
Two- and three-point functions in Wess-Zumino-Witten theories
were discussed in \cite{Ras2} and \cite{Ras3}, respectively.
They are essentially constructed by working out a basis 
$\{R^{\la,\mu,\nu}_s\}$ of the associated tensor products.
Their level-dependence was considered in \cite{RW5}.

\section{Elementary fusions}
 
When discussing fusion above, the procedure has been to decompose into
elementary couplings, implement the syzygies, and eventually consider
linear combinations of polynomials in order to determine the multi-set
of threshold levels. Here we advocate an alternative approach where 
extra EFs are introduced \cite{BKMW,RW5}. 
That will increase the number of syzygies and the complexity
of their implementation, but result in a good basis $\{R^{\la,\mu,\nu}_s\}$
for discussing fusion. Thus, the merit is that one may avoid the
final step of considering linear combinations of polynomials.
The multi-set of threshold levels is simply given by the threshold
levels of the basis polynomials: $\{t\}=\{t(R^{\la,\mu,\nu}_s)\}$.
It is natural to conjecture that only a finite number of such extra
{\em purely affine} EFs will be needed. 

Implicit in our discussion has been that all elementary couplings
are also elementary fusions. One only needs to assign their threshold
levels. However, that is easy to show using our polynomial
correspondence. We also note that any coupling $(\la,\mu,\nu)$
with $T_{\la,\mu,\nu}=1$ has a unique (up to normalization) polynomial
realization $R^{\la,\mu,\nu}$. The associated threshold level is
therefore given unambiguously by $t(R^{\la,\mu,\nu})$.
We stress, though, that not all elementary couplings correspond to tensor 
products of unit multiplicity. $G_2$ provides a simple example
\cite{GS}: $T_{2\Lambda^2,2\Lambda^2,\Lambda^1+\Lambda^2}=2$, and 
one of the two couplings (both with $t=3$) is elementary.

It is evident that the only coupling of threshold level 0 is
$(0,0,0)$. Thus, all other couplings have positive threshold levels.

\subsection{$su(r+1)$ fusion: polynomials}

In the following we will concentrate on $su(r+1)$. We recall that
${\La^i}^+=\La^{r+1-i}$ for all $i=1,...,r$. Since $\theta=\La^1+\La^r$,
we shall pay particular attention to the two fundamental highest weight
modules $M_{\La^1}$ and $M_{\La^r}$. All weights in the modules
appear with multiplicity one, so choosing a basis is straightforward.
We have made the following simple choice of bases, and listed their
associated polynomials:
\ben
 \ket{\La^1}\leftrightarrow 1,\ 
 F_{\al_1}\ket{\La^1}\leftrightarrow P_{-\al_1}^1,\ 
\ldots\ ,\ 
 F_{\al_{1,j}}\ket{\La^1}\leftrightarrow P_{-\al_{1,j}}^1,\ 
  \ldots\ ,\ 
 F_{\theta}\ket{\La^1}\leftrightarrow P_{-\theta}^1
\label{La1}
\een
and
\bea
 \ket{\La^r}\leftrightarrow 1,\ 
 F_{\al_r}\ket{\La^r}\leftrightarrow P_{-\al_r}^r,\ 
  \ldots\ ,\ 
 F_{\al_{i,r}}\ket{\La^r}\leftrightarrow P_{-\al_{i,r}}^r,\ 
  \ldots\ ,\ 
 F_{\theta}\ket{\La^r}\leftrightarrow P_{-\theta}^r
\label{Lar}
\eea
Here we have labeled the positive roots as
\ben
 \al_{i,j}=\al_i+...+\al_j\ ,\ \ \ \ \ 1\leq i\leq j\leq r
\een

Using the commutators
\ben
 [E_{\al_{i,j}},E_{\al_{k,l}}]=\delta_{k-j,1}E_{\al_{i,l}}-\delta_{i-l,1}
  E_{\al_{k,j}}
\label{EE}
\een
and the well-known symmetries of the $su(N)$ structure constants such as 
${f_{-\al,-\beta}}^{-(\al+\beta)}=-{f_{\al,\beta}}^{\al+\beta}$ and
${f_{\beta,-(\al+\beta)}}^{-\al}={f_{\al,\beta}}^{\al+\beta}$,
it is straightforward to verify that the elementary polynomial associated
to $(\La^1,\La^r,0)$ is
\bea
 R^{\La^1,\La^r,0}(x_1,x_2,x_3)&=&P^r_{-\theta}(x_2)+
  \sum_{i=1}^{r-1}P_{-\al_{1,i}}^1(x_1)P_{-\al_{i+1,r}}^r(x_2)
  -P_{-\theta}^1(x_1)\nn
 &=&x_2^\theta-x_1^\theta+...
\label{R1r}
\eea
The dots indicate a non-vanishing polynomial in $x$ independent
of $x^\theta$. One may actually derive the $x^\theta$-dependence (\ref{R1r})
without working out the entire two-point polynomial. One simply observes
that wt$(R^{\La^1,\La^r,0})=\theta$, and then considers the vanishing action 
of $E_\theta$ on $R^{\La^1,\La^r,0}$. The apparent asymmetry in signs in 
the general expression (\ref{R1r}) is due to our choices (\ref{La1}), 
(\ref{Lar}) and (\ref{EE}).
 
Useful in the following are the tensor products
\bea
 M_{\La^1}\otimes M_{\La^1}&=&M_{\La^2}\oplus M_{2\La^1}\nn
 M_{\La^1}\otimes M_{\La^r}&=&M_{0}\oplus M_{\theta}\nn
 M_{\La^r}\otimes M_{\La^r}&=&M_{\La^{r-1}}\oplus M_{2\La^r}
\label{LaLa}
\eea
which are true for $r\geq2$. 

We are now in a position to consider the triple product $M_\theta\otimes 
M_\theta\otimes M_\theta$, cf. (\ref{sun}). It is easily seen that the 
coupling $(\theta,\theta,\theta)$ admits the following two decompositions 
on the set of two-point couplings:
\bea
 (\theta,\theta,\theta)&=&(\La^1,\La^r,0)\times(0,\La^1,\La^r)\times
  (\La^r,0,\La^1)\nn
 &=&(\La^r,\La^1,0)\times(0,\La^r,\La^1)\times(\La^1,0,\La^r)
\label{ttt}
\eea
They are only distinct when $r\geq2$, which we will assume in the
following. For $r\geq3$, (\ref{ttt}) exhaust the possible decompositions.
That follows from a simple inspection of (\ref{LaLa}). For $r=2$, on
the other hand, $(\La^1,\La^1,\La^1)$ and
$(\La^r,\La^r,\La^r)$ are also elementary couplings, leading to
the well-known $su(3)$ syzygy. Summarizing the $su(3)$ case,
$T_{\theta,\theta,\theta}=2$ and the multi-set of threshold levels
is $\{2,3\}$.

For $r\geq3$, the situation is radically different. So far we have found
that $1\leq T_{\theta,\theta,\theta}\leq2$, and the two candidate
polynomials are
\bea
 R_1^{\theta,\theta,\theta}(x_1,x_2,x_3)&=&R^{\La^1,\La^r,0}R^{0,\La^1,\La^r}
  R^{\La^r,0,\La^1}\nn
 &=&(x_2^\theta-x_1^\theta)(x_3^\theta-x_2^\theta)
   (x_1^\theta-x_3^\theta)+ S_1^{\theta,\theta,\theta}(x_1,x_2,x_3)\nn
 R_2^{\theta,\theta,\theta}(x_1,x_2,x_3)&=&R^{\La^r,\La^1,0}R^{0,\La^r,\La^1}
  R^{\La^1,0,\La^r}\nn
 &=&(x_1^\theta-x_2^\theta)(x_2^\theta-x_3^\theta)
   (x_3^\theta-x_1^\theta)+ S_2^{\theta,\theta,\theta}(x_1,x_2,x_3)
\label{S}
\eea
According to (\ref{R1r}), $S_1^{\theta,\theta,\theta}$ and
$S_2^{\theta,\theta,\theta}$ are non-vanishing, second-order polynomials
in $x^\theta$. Now, if $T_{\theta,\theta,\theta}=1$ there would have to 
be a syzygy relating $R_1^{\theta,\theta,\theta}$ and 
$R_2^{\theta,\theta,\theta}$. Following (\ref{S}), that
would imply that $R_1^{\theta,\theta,\theta}+R_2^{\theta,\theta,\theta}=0$,
and in particular $S_1^{\theta,\theta,\theta}+S_2^{\theta,\theta,\theta}=0$.
The contributions to $S_1^{\theta,\theta,\theta}$ and 
$S_2^{\theta,\theta,\theta}$ that are quadratic in $x_3^\theta$ and otherwise
only depend on $x_2$, are
\ben
 S_1^{\theta,\theta,\theta}:\ \ 
  (x_3^\theta)^2(-P^r_{-\theta}(x_2)+x_2^\theta)\ ,\ \ \ \ \ \ 
 S_2^{\theta,\theta,\theta}:\ \ 
  (x_3^\theta)^2(P^1_{-\theta}(x_2)-x_2^\theta)
\een
So a syzygy would require $P^1_{-\theta}(x_2)=P^r_{-\theta}(x_2)$. 
However, it is seen (\ref{EP}) that
\bea
 P^1_{-\theta}(x)&=&x^\theta+\hf x^{\al_1}x^{\al_{2,r}}+...\nn
 P^r_{-\theta}(x)&=&x^\theta-\hf x^{\al_1}x^{\al_{2,r}}+...
\eea
Thus, we may conclude that $T_{\theta,\theta,\theta}=2$, and that
$S_1^{\theta,\theta,\theta}+S_2^{\theta,\theta,\theta}\neq0$. 

To determine the threshold levels, we must consider all non-vanishing linear
combinations of $R_1^{\theta,\theta,\theta}$ and 
$R_2^{\theta,\theta,\theta}$ to obtain a basis with lowest possible
powers of $x^\theta$. That is easily done, and we find
\ben
 R_+^{\theta,\theta,\theta}=R_1^{\theta,\theta,\theta}+
    R_2^{\theta,\theta,\theta}
\label{R+}
\een
in addition to $R_1^{\theta,\theta,\theta}$, for example. 
The threshold levels are
\ben
 t(R_+^{\theta,\theta,\theta})=2\ ,\ \ \ \ \ t(R_1^{\theta,\theta,\theta})=3
\label{tR+}
\een
$R_+^{\theta,\theta,\theta}$ does not enjoy a factorization in terms of
elementary polynomials associated to elementary couplings. Furthermore,
since it is the unique polynomial with threshold level 2, a good fusion
basis $\{R_s^{\theta,\theta,\theta}\}$ must contain it. Thus, 
$R_+^{\theta,\theta,\theta}$ corresponds to a purely affine EF.
This completes the proof of our assertion (\ref{sun}), and the explicit
construction of the associated polynomial.

When $R_+^{\theta,\theta,\theta}$ is added to the elementary couplings
in the basis of EFs, we simultaneously introduce the syzygy
\ben
 R_+^{\theta,\theta,\theta}=R^{\La^1,\La^r,0}R^{0,\La^1,\La^r}
  R^{\La^r,0,\La^1}+R^{\La^r,\La^1,0}R^{0,\La^r,\La^1}
  R^{\La^1,0,\La^r}
\een
It may be implemented by excluding the product 
$R^{\La^r,\La^1,0}R^{0,\La^r,\La^1}R^{\La^1,0,\La^r}$, for example.
That was our choice above (\ref{tR+}).

\subsection{$su(r+1)$ fusion: counting}

Here we shall deduce the existence of the purely affine elementary 
coupling (\ref{sun}) without constructing its polynomial (\ref{R+}), 
simply by calculating certain fusion multiplicities. 

A simple analysis (using the Kac-Walton formula, or modified 
Weyl character method, for example) reveals that for $r\geq3$
\ben
 \theta\otimes\theta=(0)_2\oplus (\La^2+\La^{r-1})_2
  \oplus (\La^2+2\La^r)_3\oplus (2\La^1+\La^{r-1})_3\oplus 
  (\theta)_2\oplus(\theta)_3\oplus(2\theta)_4
\label{adj}
\een
Here we have abbreviated $M_\la$ by $\la$.
The non-trivial part is the determination of the threshold levels
which have been indicated by subscripts: $(\la)_t$ corresponds
to $(\theta,\theta,\la^+)_t$, where $\la^+$ is the weight conjugate to $\la$. 
So, there exist two couplings $(\theta,\theta,\theta)$, and 
corresponding fusions with threshold levels 2 and 3. 

We need to show that (i) the two couplings $(\theta,\theta,\theta)$ 
are not elementary as couplings, and (ii) for $r>2$
the fusion $(\theta,\theta,\theta)_2$ is elementary as a fusion. 

It is not hard to show 
that {\it all} couplings corresponding to fusions at level 1 are of the form 
\ben
 \{\ (\La^i,\La^j,\La^k)_1\ |\ i,j,k\in\{0,1,2,\ldots,r\},\ 
i+j+k\in\{r+1,2(r+1)\}\ \}
\label{tec}
\een 
Here we use $\Lambda^0=0$, so that the $3r$ independent 
elementary two-point couplings are included. Clearly, each elementary 
coupling in (\ref{tec}) corresponds to an elementary fusion with threshold 
level 1.  

Now, (i) can be demonstrated easily: the two couplings 
$(\theta,\theta,\theta)$ are expressible as two distinct products of three 
couplings in (\ref{tec}), as already discussed (\ref{ttt}). 
Part (ii) is also simple: a decomposition of (\ref{sun}) 
\ben
 (\theta,\theta,\theta)_2 = 
  (\La^1+\La^r,\La^1+\La^r,\La^1+\La^r)_2=(\La^i,\La^j,\La^k)_1
  \times(\La^l,\La^m,\La^n)_1
\een
can not be on EFs; for $r>2$ there is no solution for
$\{i,j,k,l,n,m\}$ respecting $i+j+k,l+m+n\in\{r+1,2(r+1)\}$. 

We conclude by observing that for all simple Lie algebras besides
$A_r\simeq su(r+1)$, the triple adjoint couplings have tensor
product multiplicity 1 and threshold level 3. Because the adjoint
representation is a fundamental representation for all simple Lie algebras
but $A_r$ and $C_r$, the triple adjoint coupling is both an elementary 
coupling and an EF, for those algebras. For the case $C_r$, where
the highest root is $\theta=2\La^1$, it is easy to see that the triple
adjoint is neither an elementary coupling nor an EF.
In conclusion, it is only in the case $A_r$ that we encounter
a purely affine EF when considering the triple adjoint coupling.

\vskip.5cm
\noindent{\em Acknowledgements}
\vskip.1cm
\noindent J.R. is grateful to CRM, Montreal, for its generous hospitality, 
and thanks P. Mathieu for encouragement. 
He is supported in part by a PIMS Postdoctoral Fellowship and by NSERC.
M.A.W. is supported in part by NSERC, and he thanks T. Gannon for a 
discussion of the status of the threshold level conjecture.

\end{document}